\documentclass[prl,twocolumn,letterpaper, superscriptaddress,longbibliography]{revtex4-1}
\usepackage{graphicx}
\usepackage{dcolumn}
\usepackage{bm}
\usepackage{color}
\usepackage{tabularx}
\usepackage{array}

\usepackage[utf8]{inputenc}
\usepackage[T1]{fontenc}
\usepackage{color,soul}
\usepackage{tabularx}
\usepackage{amsmath}

\makeatother
\begin{document}
\title{Magneto-optical spectroscopy based on pump-probe strobe light} 

\author{Shihao Zhou}
\affiliation{Department of Physics and Astronomy, University of North Carolina at Chapel Hill, Chapel Hill, NC 27599, USA}

\author{Yujie Zhu}
\affiliation{Department of Materials Science and Engineering, University of Wisconsin-Madison, Madison, Wisconsin, 53706, USA}

\author{Chunli Tang}
\affiliation{Department of Electrical and Computer Engineering, Auburn University, Auburn, Alabama 36849, USA}
\affiliation{Department of Physics, Auburn University, Auburn, Alabama 36849, USA}

\author{Rui Sun}
\affiliation{Department of Physics and Organic and Carbon Electronics Lab (ORaCEL), North Carolina State University, Raleigh, NC 27695, USA}

\author{Junming Wu}
\affiliation{Department of Physics and Astronomy, University of North Carolina at Chapel Hill, Chapel Hill, NC 27599, USA}

\author{Yuzan Xiong}
\affiliation{Department of Physics and Astronomy, University of North Carolina at Chapel Hill, Chapel Hill, NC 27599, USA}
\affiliation{Department of Electrical and Computer Engineering, North Carolina A$\&$T State University, Greensboro, NC 27411, USA}

\author{Ingrid E. Russell}
\affiliation{Department of Physics and Astronomy, University of North Carolina at Chapel Hill, Chapel Hill, NC 27599, USA}

\author{Yi Li}
\affiliation{Materials Science Division, Argonne National Laboratory, Argonne, IL 60439, USA}

\author{Dali Sun}
\affiliation{Department of Physics and Organic and Carbon Electronics Lab (ORaCEL), North Carolina State University, Raleigh, NC 27695, USA}

\author{Frank Tsui}
\affiliation{Department of Physics and Astronomy, University of North Carolina at Chapel Hill, Chapel Hill, NC 27599, USA}

\author{Binbin Yang}
\affiliation{Department of Electrical and Computer Engineering, North Carolina A$\&$T State University, Greensboro, NC 27411, USA}

\author{Valentine Novosad} 
\affiliation{Materials Science Division, Argonne National Laboratory, Argonne, IL 60439, USA}

\author{Jia-Mian Hu}
\affiliation{Department of Materials Science and Engineering, University of Wisconsin-Madison, Madison, Wisconsin, 53706, USA}

\author{Wencan Jin}
\affiliation{Department of Physics, Auburn University, Auburn, Alabama 36849, USA}
\affiliation{Department of Electrical and Computer Engineering, Auburn University, Auburn, Alabama 36849, USA}

\author{Wei Zhang}
\thanks{Corresponding to: zhwei@unc.edu}
\affiliation{Department of Physics and Astronomy, University of North Carolina at Chapel Hill, Chapel Hill, NC 27599, USA}

\begin{abstract}

We demonstrate a pump-probe strobe light spectroscopy for sensitive detection of magneto-optical dynamics in the context of hybrid magnonics. The technique uses a combinatorial microwave-optical pump-probe scheme, leveraging both the high-energy resolution of microwaves and the high-efficiency detection using optical photons. In contrast to conventional stroboscopy using a continuous-wave light, we apply microwave and optical pulses with varying pulse widths, and demonstrate magneto-optical detection of magnetization dynamics in Y$_3$Fe$_5$O$_{12}$ films. The detected magneto-optical
signals strongly depend on the characteristics of both the microwave and the optical pulses as well as their relative time delays. We show that good magneto-optical sensitivity and coherent stroboscopic character are maintained even at a microwave pump pulse of 1.5 ns and an optical probe pulse of 80 ps, under a 7 megahertz clock rate, corresponding to a pump-probe footprint of $\sim 1\%$ in one detection cycle. Our results show that time-dependent strobe light measurement of magnetization dynamics can be achieved in the gigahertz frequency range under a pump-probe detection scheme.

\end{abstract}

\flushbottom
\maketitle

\thispagestyle{empty}

\section{I. Introduction}

Pump-probe spectroscopy and microscopy experiments represent an important branch of tools to study the electronic, thermal, and spin dynamics in novel materials and their associated non-equilibrium states \cite{fischer2016invited,lee2020high}. In this technique, an ultrashort pump pulse excites the sample, generating a non-equilibrium state, and a weaker, probe pulse monitors the pump-induced changes in the optical characteristics of the sample. In most pump-probe setups, the pump and probe pulses are both optical in nature. However, combinatorial-wavelength systems also exist, where the pump and probe signals are at different energy scales \textcolor{black}{across microwave \cite{tamaru2002imaging}, terahertz \cite{dastrup2024optical}, and x-rays \cite{burn2020depth}}. One prevalent example of such is the optically detected magnetic resonance (ODMR) \cite{clevenson2015broadband,rondin2014magnetometry,mittiga2018imaging,zhou2024sensing}, in which a microwave (MW) pump is used in lieu of an optical one, allowing one to benefit from the high energy resolution of the microwaves (down to a few 100 kHz, i.e. neV energy range) with the high efficiency of optical photon detection. 

Separately, stroboscopy represents another class of techniques, in which a strobe light, synchronized with the fundamental frequency of interest, flashes at a slightly slower (or faster) rate to capture successive phases of the oscillatory cycle \cite{d2024stroboscopic,whiteley2019correlating,hinterstein2023stroboscopic}. In this case, a continuous-wave (cw) light is commonly used, which, after a periodic modulation of its intensity, allows one to accurately snap-capture, and hence, trace the dynamics \cite{nembach2013mode,yoon2016phase}, with explicitly defined phase contributions by a ``slow motion effect''. 

While the strobe light nicely captures the phase information of the dynamics due to coherent modulation, the cw nature of the light emission may potentially result in nontrivial, local heat buildup or thermal damage, rendering the cw format unfavorable in a number of measurement situations, e.g., where the target material is sensitive to heat or when high energy densities are involved. In addition, the precise temporal evolution of the non-equilibrium state may still be of interest apart from the phase information. Therefore, it is tempting to combine the pump-probe scheme with the strobe light character to retain a good phase resolution while significantly reducing the input energy footprint, e.g. approaching ``single-shot'' optical pulse detection \cite{gunyho2024single,gritsch2025optical,arnold2025all,song2025single}. However, such a system has remained elusive.    

\begin{figure*}[htb]
 \centering
 \includegraphics[width=6.6 in]{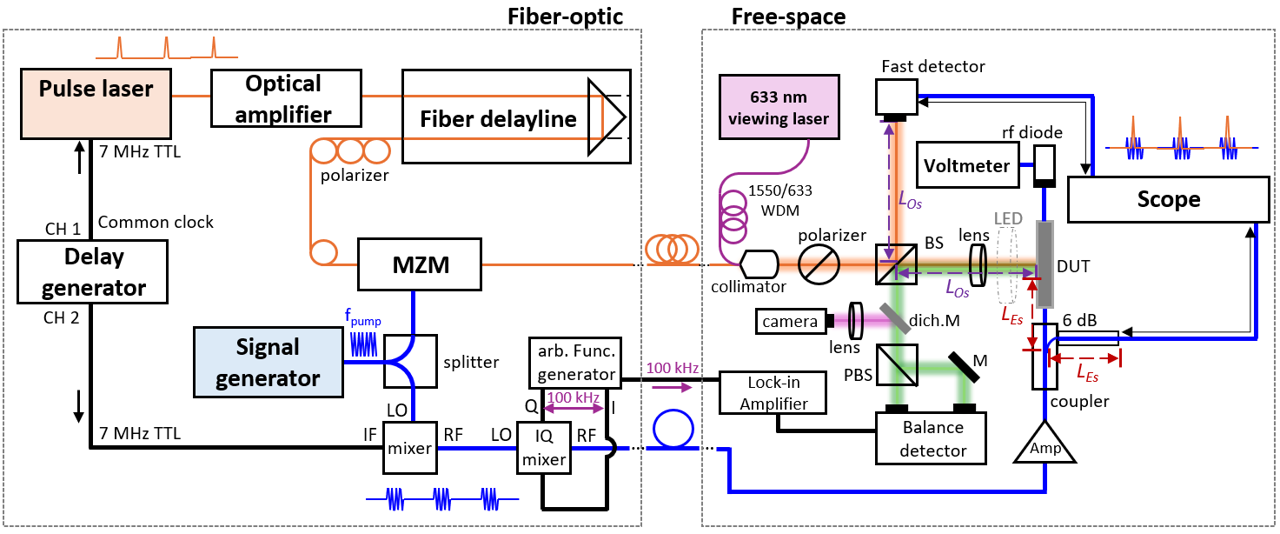} 
 \caption{Schematic illustration of the optoelectronic setup. (Left) Fiber-optic panel: the delay generator synchronizes the laser pulse and the \textcolor{black}{microwave (MW)} pulse controlled by distinct but locked channels, CH1 and CH2. The optical pulse train, after passing through the optical amplifier, the fiber delayline, and the polarizer, is intensity-modulated at the \textcolor{black}{Mach-Zehnder Modulator (MZM)}, at a frequency of the MW signal (upper branch after the rf splitter). The MW pulse train is generated from the same MW signal by a rf mixer (lower branch after the splitter). The MW pulse width is controlled by the delay generator. This signal is then \textcolor{black}{in-phase and quadrature (I-Q)} mixed to enable a heterodyne detection mechanism. The I and Q signals \textcolor{black}{(100 kHz in the present work)} are sourced by an arbitrary function generator with appropriate voltage level and phase lag. (Right) Free-space panel: after the fiber collimator, the light pulses are polarized at 45$^\circ$, passing through the beam splitter, and reaching the sample \textcolor{black}{device-under-test (DUT)}. The magnetization dynamics of the DUT induces a polarization modulation that can be sensitively captured -- with the birefringence effect -- using a \textcolor{black}{polarization beam splitter (PBS)}, a balanced photodetector, and the lock-in amplifier. The excitation MW pulses are amplified, passing through a 6-dB coupler, and reaching the sample DUT. To monitor the relative time delay between light and MW pulses arriving at the DUT, two sampling branches are installed: optically, from the \textcolor{black}{beam splitter (BS)} to an ultrafast detector (optical sampling path, $L_\mathrm{Os}$), and electrically, at the coupler end (electrical sampling path, $L_\mathrm{Es}$). Additionally, the laser alignment and sample viewing is achieved via a visible light path, consisting of a red laser, a 1550/635 \textcolor{black}{wavelength division multiplexing (WDM)}, a dichroic mirror, and a CMOS camera. Between the two panels, fiber-optic and microwave patch cables with appropriate lengths are inserted to compensate for any outstanding path differences. \textcolor{black}{IF: intermediate frequency (mixer), LO: local-oscillator (mixer), RF: radio-frequency (mixer), arb.Func.generator: arbitrary functional generator, Amp: Amplifier, M: mirror, dich.M: dichroic mirror. } }  
 \label{fig_scheme}
\end{figure*}

Here, we demonstrate a pump-probe strobe light spectroscopy for detecting magneto-optical (MO) dynamics in the context of hybrid magnonics \cite{awschalom2021quantum,lachance2019hybrid,li2020hybrid,yuan2022quantum}. The study of magnons, i.e. the fundamental excitations of magnetic materials \cite{chumak2022advances,flebus20242024}, manifests a superior testbed for optical spectroscopic techniques: \textit{one}, the magnetic birefringence, through the magneto-optical Faraday, Voigt (Cotton–Mouton), and Kerr effects \cite{freiser1968survey,kimel20222022}, remains one of the most widely applied phenomena in modern photonics, broadly extending from UV to IR regimes; \textit{two}, hybrid microwave-optical pump-probe method can be used as in the ODMR, leveraging coherent excitations using mature developments in microwave-magnetics (waveguide, resonator, etc) \cite{bhoi2019photon,harder2018cavity}; \textit{three}, the measured spectral, temporal, and phase characteristics directly link to the critical physical properties of magnons, such as dispersion, propagation, and nonlinearity, which have been found useful in magnon-based information technologies \cite{chumak2015magnon,wang2024nanoscale}.     

\textcolor{black}{In a typical pump-probe spectroscopy, both the pump and probe (whether electrical or optical) manifest as trains of identical pulses with a single pulse form, without any further intensity modulation (and hence any frequency specificity) \cite{fischer2016invited,lee2020high}. In contrast, our technique adopts a MW (pump) and optical (probe) pulse train} that are frequency-specific \textcolor{black}{due to an additional intensity-modulation on the pulse series}, so that the coherent strobe feature is preserved \textcolor{black}{in a time-averaged detection}.  On the other hand, in contrast to prior works where both the MW (pump) and optical light (probe) are in cw form \cite{xiong2020detecting,xiong2020probing,xiong2022tunable,xiong2024phase,christy2025tuning}, we use MW and optical pulses with varying pulse widths and demonstrate MO detection of magnon dynamics in Y$_3$Fe$_5$O$_{12}$(YIG) films \cite{serga2010yig}. The detected magneto-optical signals strongly depend on the characteristics of both the MW and the optical pulses, as well as their relative time positions (delays). In particular, we show that good MO sensitivity and coherent strobe light character can still be maintained even down to a MW pump pulse of 1.5 ns and an optical probe pulse of 80 ps under a 7 MHz clock rate, corresponding to a pump-probe footprint of $\sim 1\%$ in one detection cycle. Our results show that time-dependent strobe light measurement can be achieved in the gigahertz (GHz) frequency range under a pump-probe detection scheme.

\section{II. Experiments}

\subsection{A. Optical Setup}

Our experimental setup is illustrated in Fig.\ref{fig_scheme}. Leveraging the rich library of fiber-optic components available in the telecom band, a good amount of the optical layout can be constructed in an all-fiber format (left panel), \textcolor{black}{and the costly femtosecond (fs) mode-locked laser (that is often encountered in typical pump-probe setups) can be avoided}  \cite{xiong2024magnon,wu2025coupling}. Free-space optics (right panel) are only used close to the sample end. 

In the fiber-optic panel, instead of using a cw telecom laser, we use a gain-switched picosecond pulsed laser (Thorlabs GSL155A) operating at 1550-nm for the demonstration, in combination with the use of microwave pulses for exciting the magnetization dynamics (of the DUT). The laser pulse width can be conveniently tuned from $<80$ ps to 65 ns, allowing to examine a ``pulse width''-dependence on the MO signal. The microwave signal is provided by a microwave generator (Berkeley Nucleonics BNC845) in cw form, and the pulsed microwave is generated using an ultrafast rf mixer. A digital delay generator (Stanford Research DG645) serves as the common time clock for synchronizing the laser and the microwave pulses (via the IF port of the mixer). We use a clock frequency of 7 MHz, corresponding to a time period $T = 142.9$ ns. Two separate channels, CH1 and CH2, are used for the laser and microwave respectively, whose relative delay can be tuned digitally in the delay generator up to $25$ ns for each channel (coarse delay). 

Along the optical path, a unique addition is a fiber-optic delayline (Chongqing Smart Science $\&$ Technology, up to 5000 ps delay) inserted between the optical amplifer (Thorlabs EDFA100S) and the fiber polarization controller (Thorlabs FPC032), in contrast to the previous configuration using cw light. The fiber-optic delayline offers finer control of the delay time than the digital delay set by the DG645. Along the electrical path, the cw microwave signal generated by the BNC845 is first divided by an rf splitter, one branch (retaining its cw form) is used to modulate the pulsed strobe light at the Mach-Zehnder Modulator (MZM); the other branch (pulsed by the mixer) is used to excite the magnetization dynamics of the DUT. An in-phase/quadrature(I/Q) mixer is installed to \textcolor{black}{enable a sideband generation} and heterodyne signal detection using a lock-in amplifier (Stanford Research SR830), in which the I and Q ports are sourced by an arbitrary function generator, providing 1.0 Vpp at 100.0 kHz, with 67$^\circ$ I-to-Q phase lag \textcolor{black}{for optimizing the higher sideband, so that its power is >20 dB higher than the central frequency and the lower sideband.}  

In the free-space regime, the optical layout generally retains the same configuration as in the cw light case: a polarization modulation (by the sample DUT) to an initial 45$^\circ$-polarized light is sensitively captured -- with the MO birefringence effect -- by using a polarization beam splitter (PBS), a balanced photodetector, and the lock-in amplifier. Using a wavelength division multiplexer(WDM) that combines 1550/635-nm (Seagnol Photonics), a red fiber-laser (633-nm), combined with a ring-LED, a long-pass dichroic mirror, and a CMOS camera, can be integrated for easy alignment of laser spot and sample viewing. 

To accommodate the use of light and microwave pulses, two sampling branches are uniquely installed for monitoring the travel of the respective optical and microwave pulses. At the center beam splitter (BS), the previously unused, deflected portion of light is recycled and made to travel the same distance as that to the DUT, denoted as the optical sampling, $L_\mathrm{Os}$, before it is received by a high-speed optical detector (Thorlabs DET08CL). Similarly, the microwave sampling is achieved by using a 6-dB directional coupler (Pasternack PE2CP004-6), in which the split microwave is made to travel the same distance as that to the DUT, denoted as the electric sampling, $L_\mathrm{Es}$. The sampling pulses are then acquired and monitored by a high-speed oscilloscope (Pico Technology PicoScope 9300). Finally, additional fiber-optic and microwave patch cables with appropriate lengths are inserted between the two panels to compensate for any outstanding path differences.

\begin{figure}[htb]
 \centering
 \includegraphics[width=3.1 in]{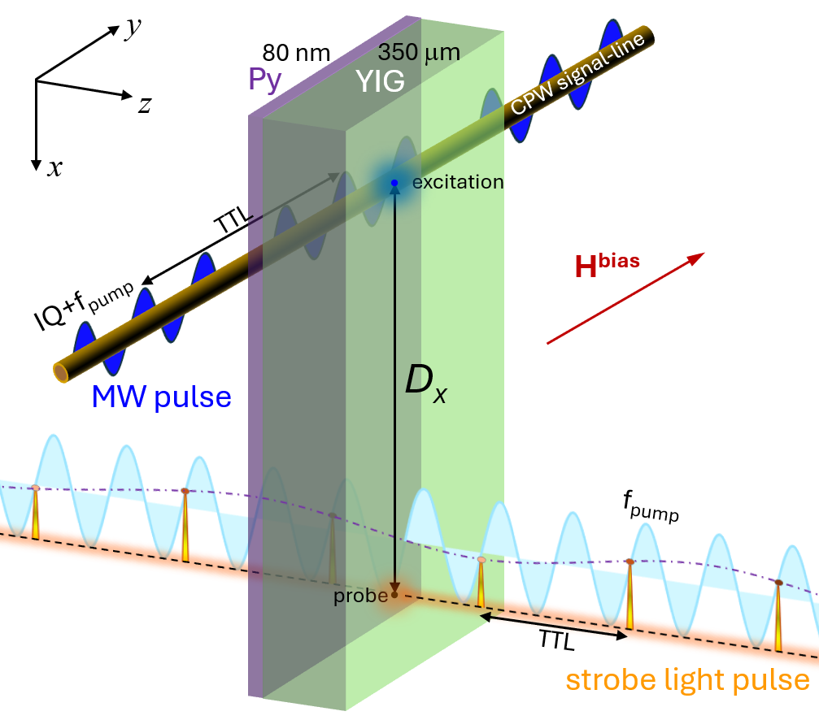}
 \caption{Sample and measurement scheme: \textcolor{black}{the sample DUT consists of a YIG(350-$\mu$m)/Py(80-nm) bilayer sample centered atop a CPW under the flip-chip configuration (Py facing CPW). An external bias magnetic field is applied along the $y$ direction ($H_y^\textrm{bias}$), and the oscillation rf field is along $x$. The optical pulse train is intensity-modulated at $f_\textrm{pump}$, and the MW pulse train, at \textcolor{black}{IQ(100 kHz)}$+f_\textrm{pump}$, is delivered via the CPW signal-line, so that the coherent strobe feature is maintained in the measurement. The optical pulse train probes at a sample location that is a distance ($D_x$) away from the CPW's signal-line.} }   
 \label{fig_demo}
\end{figure}

\subsection{B. Sample and Measurements}

\begin{figure*}[htb]
 \centering
 \includegraphics[width=7.3 in]{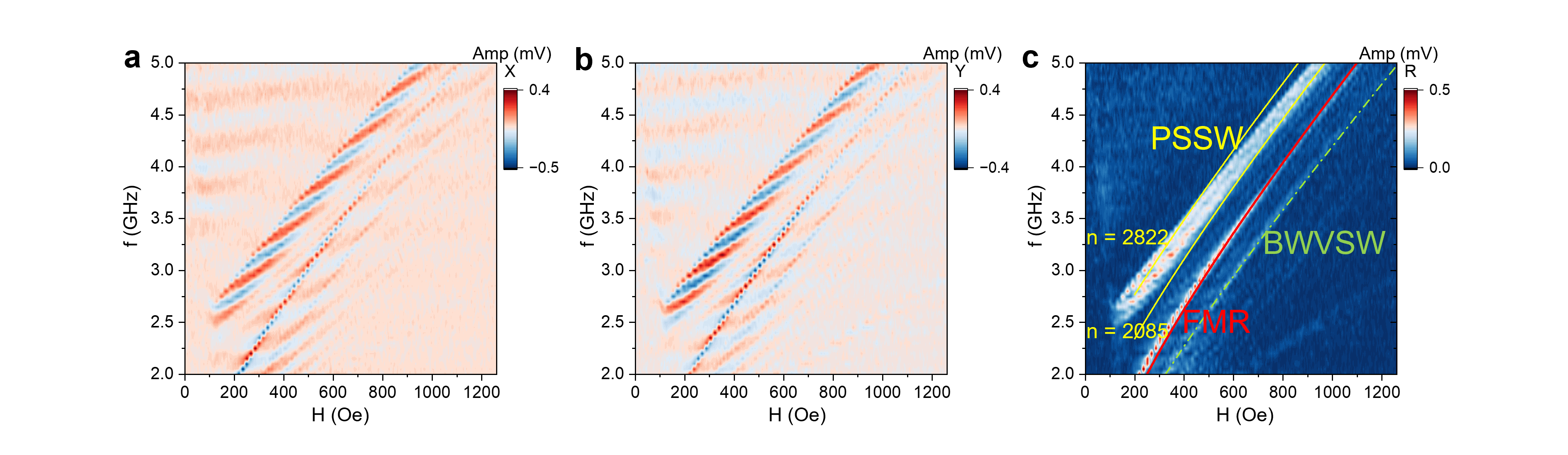}
 \caption{Pulse strobe light detection of magnon dispersion. \textcolor{black}{(a-c)} Magnon dispersion [$f,H$]-contour plots from the lock-in amplifier's (a) $X$, (b) $Y$, and (c) $R$ channels, measured from 2 -- 5 GHz. The MW pulse is 25 ns and the optical pulse is 20 ns. \textcolor{black}{Apart from the central FMR mode, selective PSSW and BWVSW modes are also excited. } }  
 \label{fig_dispersion}
\end{figure*}

To demonstrate the capability of the setup, we measured a YIG/Permalloy(Py) bilayer in which the Py layer of 80-nm thickness was deposited atop a commercial, double-side-polished YIG disc pallet of 350-$\mu$m thickness and a diameter of 5-mm. \textcolor{black}{The YIG disc is a single-crystal pallet along the <111> orientation, grown by the float-zone method. The mid-IR transmittance is $>70\%$. The FMR linewidth is less than 5 Oe in the measured frequency range \cite{xiong2024phase}. } The bilayer sample is placed atop a coplanar waveguide (CPW) with the Py surface facing against the CPW structure, as illustrated in Fig.\ref{fig_demo}. 

\textcolor{black}{The common clock (TTL) controls the rate of the MW pulse and the strobe light pulse. The MW and optical pulses are synchronized at the sample DUT, whose relative time position can be tuned by adjusting the delay time. In contrast to the conventional pump-probe technique, the optical probe pulse train is frequency-specific due to an additional intensity-modulation on the pulse series, at a frequency, $f_\textrm{pump}$, sourced by the signal generator, see Fig. \ref{fig_demo}. The same $f_\textrm{pump}$, after I-Q mixing, generating a sideband at \textcolor{black}{IQ(100 kHz)}$+f_\textrm{pump}$, drives the spin dynamics of the sample DUT, so that the coherent strobe feature is preserved in a time-averaged detection \cite{xiong2024phase}, even with a pulsed light form. }  

The rf excitation field, $h_\textrm{rf}$, oscillates along the $x$ direction. An external bias magnetic field is applied along the $y$ direction, i.e. $H_y^\textrm{bias}$. \textcolor{black}{The distance between the excitation point (at the CPW) and the probe point (at the laser spot) is denoted as $D_x$.} Due to the rather thick layer of both the Py and YIG, the magnon-magnon coupling \cite{xiong2020probing,xiong2022tunable,xiong2024phase,li2020coherent,qin2018exchange,liensberger2019exchange,chen2018strong} at the interface is quite weak, therefore, the Py layer merely serves as a reflective surface for the strobe light probe. All measurements occur at room temperature, 297.15 K.


\section{III. Results and Discussions}

\subsection{A. Strobe light detection of magnon spectrum}

As the YIG's magnetization, $\textbf{m}^0(x,y,z,t)=\textbf{m}^0+\Delta \textbf{m}^0(t)$, is in-plane in the off-resonance state, the film-normal component of the magnetization ($m^0_z$) is zero. Once the ferromagnetic resonance(FMR) is reached, this $m^0_z$ oscillates as a result of pumping at the MW frequency, giving rise to an oscillating $\Delta m^0_z(t)$, and hence, an instantaneous MO signal caused by the instantaneous polarization rotation of the reflected light. The optically measured rotation comes from time-averaged light intensity where the averaging is slow compared to the MW frequency, but fast compared to the \textcolor{black}{IQ signals} \textcolor{black}{injected to the mixer}. The measured MO rotation is therefore given by the time-averaged product of the rotation angle, which is proportional to $\Delta m^0_z(t) \propto m^0$ and the optical power, $P_\textrm{o}$. 

We first measure the magnon dispersion in a broad frequency range of 2--5 GHz, using a MW pulse of 25 ns paired with an optical pulse of 20 ns. To maximize the detected signal, the MW and optical pulses are synced to temporally overlap with each other, by properly adjusting the fiber and microwave delays and via monitoring their time positions from the respective sampling paths, \textcolor{black}{the $L_\mathrm{Os}$ and $L_\mathrm{Es}$ in Fig. \ref{fig_scheme}}. The stroboscopic mechanism still applies using pulsed light, as shown by the optically detected magnon dispersion presented in \textcolor{black}{Fig.\ref{fig_dispersion}(a-c)}. The essence of the detected signal after demodulation in the lock-in amplifier captures the total accumulation of the strobe phase, $\phi_\textrm{strobe}$, in which the lock-in amplifier's in-phase ($X$) channel, Fig.\ref{fig_dispersion}(a), and quadrature ($Y$) channel, Fig.\ref{fig_dispersion}(b), read: $X \propto m^0 P_\textrm{o} \textrm{cos}\phi_\textrm{strobe}$ and $Y \propto m^0 P_\textrm{o} \textrm{sin}\phi_\textrm{strobe}$, and the total MO signal, via the amplitude ($R$) channel, Fig.\ref{fig_dispersion}(c), $V_\textrm{amp} \propto m^0 P_\textrm{o}$. 

\textcolor{black}{The strobe phase constitutes two components, one due to the \textcolor{black}{electrical-optical} (E-O) path difference, $\phi_\textrm{eo} = 2\pi f_\textrm{pump} \Delta_\textrm{EO}/c$ \cite{xiong2024phase}, and another due to the magnon propagation (along $D_x$), $\phi_\textrm{mp} = 2\pi f_\textrm{pump} D_x/v_g$, in which $\Delta_\textrm{EO}$ denotes the E-O path difference, $c$ the speed of light, and $v_g$ the group velocity of the magnons. As can be seen, both phases are linearly proportional to the modulation frequency, $f_\textrm{pump}$, which is responsible for the periodic banding (red - blue) in the $f-H$ dispersions captured by the lock-in amplifier's $X$ and $Y$ channels, Fig. \ref{fig_dispersion}(a,b).} More details on the contributions to the measured strobe phase can be found in earlier reports \cite{xiong2020probing,xiong2022tunable,xiong2024phase}.

\begin{figure}[htb]
 \centering
 \includegraphics[width=3.4 in]{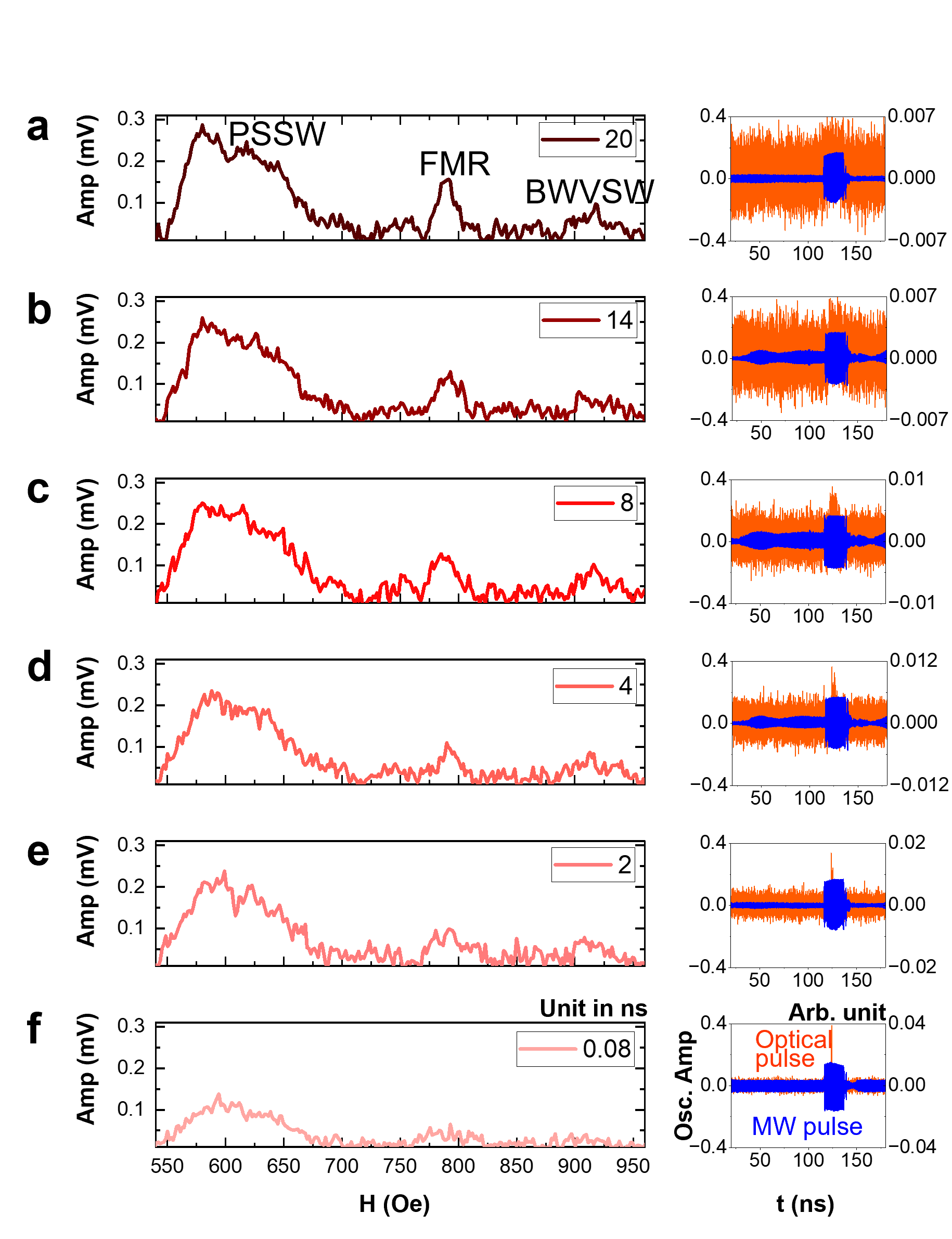}
 \caption{Optical pulse width dependence: (a-f) (Left panel) magnon spectrum measured at a fixed MW pump pulse (20 ns, 4-GHz) but with various optical probe pulse widths (unit in ns): (a) 20, (b) 14, (c) 8, (d) 4, (e) 2, and (f) 0.08. (Right panel) The relative time positions of the MW pump and the optical pulses are monitored in real-time using an oscilloscope, taking advantage of the MW and optical sampling paths of our setup.  }  
 \label{fig_opt}
\end{figure}

The measured $f-H$ magnon dispersion clearly reveals the Kittel mode of YIG, and a theoretical fitting to the dispersion \textcolor{black}{(See Appendix for the detailed derivation)} yields a saturation magnetization of YIG, $M_\textrm{s}$ of $1.409\times10^5$ A/m, using a gyromagnetic ratio $\gamma$ of $2.21\times10^5$ Hz m/A, an exchange stiffness $A_{ex}$ of $2.6\times 10^{-12}$ J/m, and an anisotropy $K_1$ of $-572.3$ J/m$^3$ \cite{chikazumi1997physics}. In addition, a pronounced magnon band is observed as a result of the pulsed MW excitation, which is different from the conventional cw scenario.  The MW pulse excites a magnetization pulse in the YIG and generates the \textcolor{black}{perpendicular standing spin-wave} (PSSW) magnon modes oscillating in the $z$ direction, which can subsequently propagate along the in-plane ($x,y$) directions and be detected by the optical probe \cite{sheng2023nonlocal}. \textcolor{black}{The PSSWs have frequencies higher than the FMR mode. In addition, the geometry also excites the backward volume spin-wave (BWVSW) modes at frequencies lower than the FMR mode, see Fig. \ref{fig_dispersion}(c), though they are not focused in the present study}.  
 
Given our excitation and detection geometry, we are interested in the in-plane magnon propagation along the $x$ direction, see Fig. \ref{fig_demo}. We assume the initial magnetization excited by the single-frequency MW pulse has a Gaussian shape, i.e., $\textbf{m}^0(x,y,z,t)=\textbf{m}^0+\Delta \textbf{m}^0 \textrm{exp} \big(-\frac{(x-x_0)^2+(y-y_0)^2+(z-z_0)^2}{2\sigma^2} \big) \textrm{exp} \big(i(k_x^0 x+k_z^0 z-\omega t) \big)$, where $x_0, y_0, z_0$ are the coordinates of the MW pump pulse center and $\sigma$ is the characteristic size of the excited initial magnetization. \textcolor{black}{This assumption is reasonable for two reasons. First, the amplitude of the excited magnetization has a 3D isotropic Gaussian distribution in the real space, which captures the intensity distribution of the excitation MW pump pulse in the YIG to a 1$^\textrm{st}$-order approximation. Second, the initial magnetization distribution in the $k$-space, which can be obtained by performing a 2D Fourier transform in the $x-$ and $z-$direction, is a Gaussian shape centered at \textcolor{black}{a specific wavenumber $k_x^0$ and $k_z^0$ }with a bandwidth $\sigma_k=1/\sigma$. This adds a constraint that the $k_z$ of the excited PSSW magnons fall within a limited range, which is consistent with the observed PSSW band in Fig. \ref{fig_dispersion}(c).}     

We analyze the PSSW band in Fig.\ref{fig_dispersion}(c) considering a cavity size ($d_\textrm{YIG}$) of 350 $\mu$m, and extract the upper and lower values of the wavevector ($n\pi/d_\textrm{YIG}$): $k_\textrm{up}=8.01\times10^{7}$ m$^{-1}$, $(n = 2822)$, and $k_\textrm{lo}=5.92\times10^{7}$ m$^{-1}$, $(n = 2085)$, and a center wavevector $k_\textrm{center}=\frac{1}{2}(k_\textrm{up}+k_\textrm{lo}) = 6.96\times10^7$ m$^{-1}$. Such a wavevector range corresponds to the full width of a Gaussian shape covering approximately $\pm8\sigma_k$, thus the standard deviation in the $k$-space is estimated as $\sigma_k = \frac{1}{16}(k_\textrm{up}-k_\textrm{lo}) = 1.3\times10^6$ m$^{-1}$. This further translates to a spatial width of the excited magnetization distribution in real space $\sigma = 1/\sigma_k = 7.63\times10^{-7}$ m ($\sim 1\mu$m). However, the actual spatial width of the excited magnetization is likely underestimated, because the $k$-range extracted from the dispersion relation is limited by the frequency resolution of the measurement and the curvature of the magnon dispersion. As a result, the true spatial extent of the excited magnetization pulse could be significantly broader, and is practically of the order of tens of $\mu$m.

\subsection{B. Optical pulse width and power dependence}

From the probe side, given that the stroboscopic mechanism still applies with pulsed light, the next question that naturally arises is, what would be the \textit{shortest optical pulse that can still be used for such a pump-probe strobe light scheme?}

Leveraging the convenient pulse shape tuning function of the gain-switched laser, we probed the MO signal at a fixed MW frequency of 4-GHz and a fixed MW pulse width of 20 ns, but using selective optical pulse widths at 0.08, 2, 4, 8, 14, and 20 ns, as shown in Fig. \ref{fig_opt}. To ensure a good overlap between the MW pump pulse and the optical probe pulse, we employed the delay components (digital or fiber) of our system, and monitored the relative position between the pump and probe by virtue of the sampling paths ($L_\mathrm{Os}$ and $L_\mathrm{Es}$). The MO signal amplitude scales almost linearly with the applied pulse width, as long as the width is smaller than the MW pump pulse. This is consistent with the fact that the measured MO signal comes from the time-averaged rotation angle, and given that the pump-probe repetition rate (7 MHz) is much higher than the \textcolor{black}{IQ-}modulation frequency (100 kHz).

\begin{figure}[htb]
 \centering
 \includegraphics[width=3.5 in]{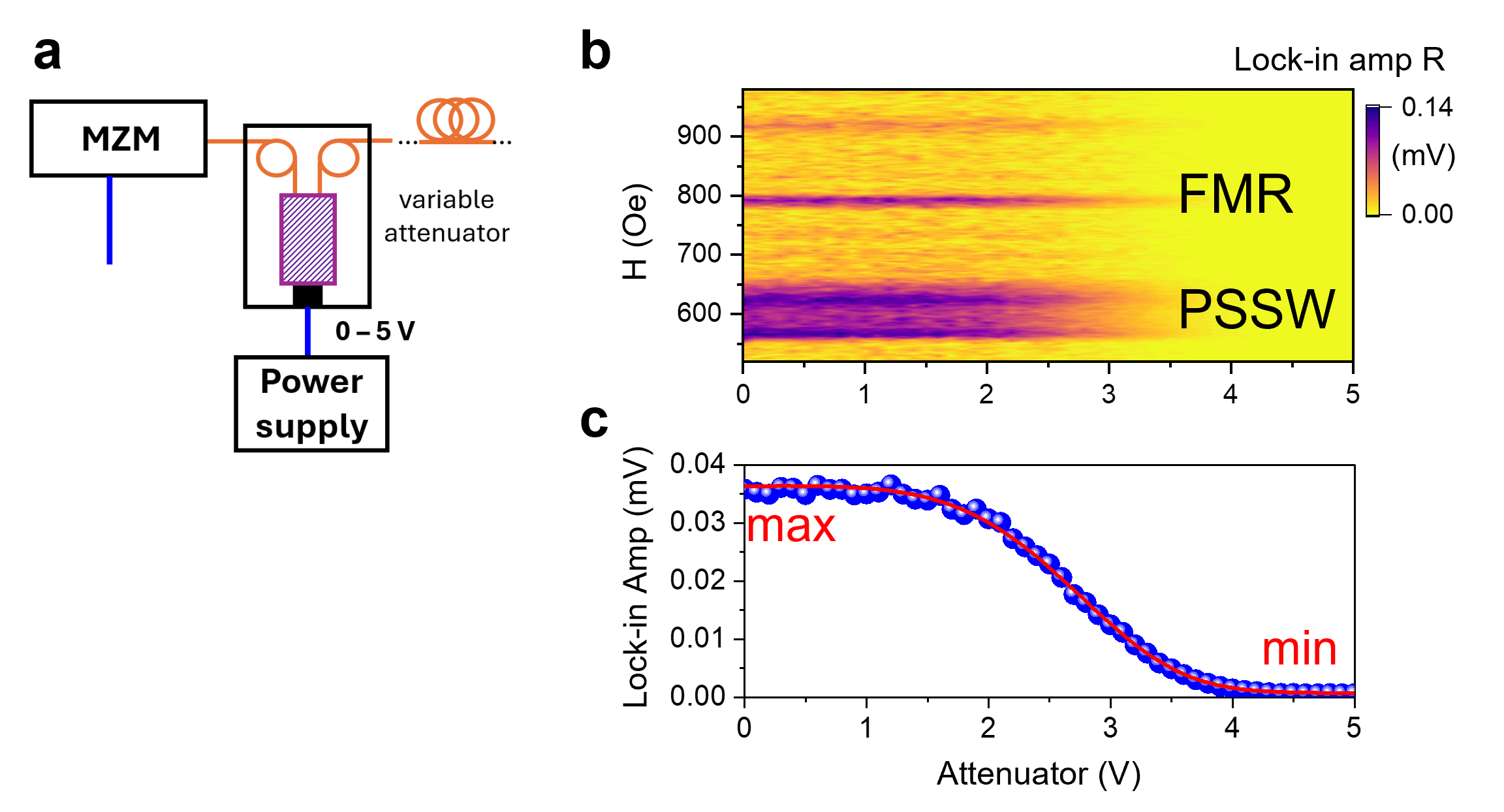}
 \caption{Optical power dependence: (a) insertion of a fiber-optic attenuator. (b) Magnon spectrum measured at different optical power attenuation levels and displayed in [$H,V$]-contour plot (0V: max optical power, 5V: min optical power), using a 25-ns MW pulse at 4 GHz, and a 12-ns optical probe pulse. (c) The total signal amplitude plotted against different attenuation levels (symbols) and fitted to the response function (curve) of the attenuator. }  
 \label{fig_opt_power}
\end{figure}

In addition, the measured MO signal corresponds to the time-averaged product of the rotation angle, and the detected voltage signal writes: $V_\textrm{amp} \propto m^0 P_\textrm{o}$, directly proportional to the optical power $P_\textrm{o}$. To validate such a power dependence, we inserted a digitally-controlled, variable optical attenuator (Thorlabs V1550A) after the MZM component in the optical path, to systematically adjust the optical power, see Fig.\ref{fig_opt_power}(a). This attenuator allows to tune the optical transmission in-fiber from $100-0\%$ (max to min) in a continuous fashion by adjusting an applied control voltage ($0-5$ V). We used an optical pulse (12 ns pulse width), synced with a MW pulse (25 ns pulse width) for the measurement. Figure \ref{fig_opt_power}(b) shows the [$H,V$]-contour plot, in which the magnetic field is scanned near the magnon resonances at a fixed frequency of 4 GHz, under different levels of optical attenuation (0V: null attention, 5V: full attenuation). Clear decreasing trend of the measured MO signal is observed. Figure \ref{fig_opt_power}(c) plots the integrated MO signal amplitude against the attenuation level (symbol), and was fitted to the response function of the attenuator (curve). The results show good agreement, indicating the linear proportionality of the MO signal to the optical power, $P_\textrm{o}$.

\subsection{C. Microwave pulse width and power dependence}

 From the pump side, the spin precessional amplitude $\Delta m^0_z(t)$ is proportional to the driving rf field, $h_\mathrm{rf}$, which relates to the MW power by, $h_\textrm{rf} = \sqrt{\frac{2P_a}{\textcolor{black}{Z_\textrm{C}}}}/d_\textrm{CPW}$ \cite{qu2025pump}, that is determined by the applied MW peak power, $P_\textrm{a}$, \textcolor{black}{the CPW impedance $Z_\textrm{C}$}, and the width of the CPW signal line, $d_\textrm{CPW}$. Thus, the detected voltage: $V_\textrm{amp} \propto m^0 P_\textrm{o} \propto \sqrt{P_\textrm{a}}P_\textrm{o}$, i.e. should follow an inverse quadratic function of $P_\textrm{a}$.  

\begin{figure}[htb]
 \centering
 \includegraphics[width=3.5 in]{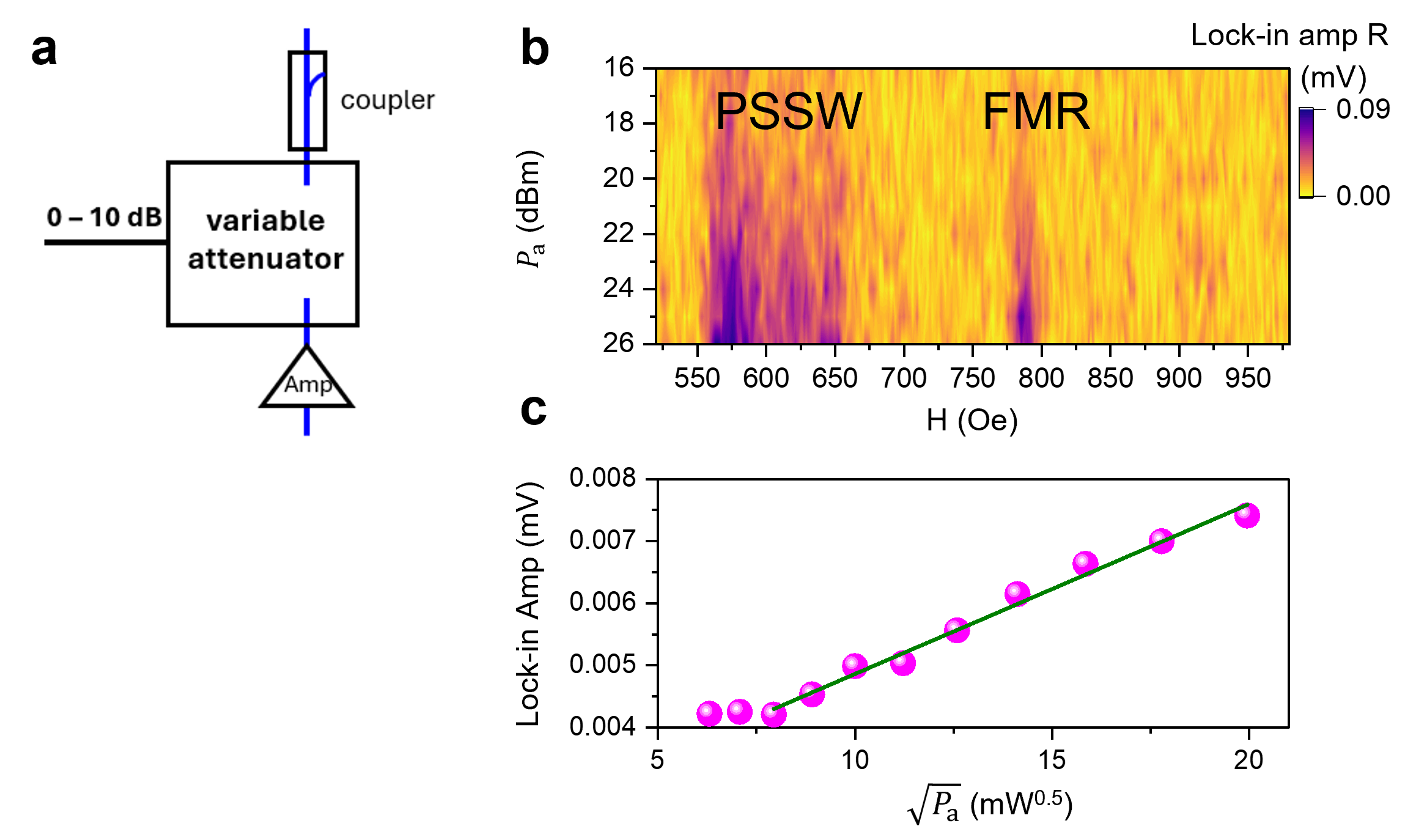}
 \caption{MW power dependence: (a) insertion of a variable MW attenuator. (b) Magnon spectrum measured at different MW power levels under controlled attenuation and displayed in a [$P_\textrm{a},H$]-contour plot, using a 25-ns MW pulse (at 4 GHz) and a 12-ns optical pulse. (c) The total signal amplitude plotted against different power levels (symbols) and linearly fitted to the theoretical model (curve) of MW power dependence. }  
 \label{fig_MW_power}
\end{figure}

To examine the MW power dependence, we inserted a programmable MW attenuator (Mini-circuits, RUDAT-8000-30) in the electrical path, after the final amplifier and before the rf coupler, see Fig. \ref{fig_MW_power}(a). The initial power arriving at the DUT, before any attenuation, is determined as 26 dBm peak power using a fast-sampling rf power meter (Mini-circuits, PWR-8PW-RC). We then reduced such a power incrementally at a step of 1 dB and measured magnon resonance, at a fixed frequency of 4 GHz.  

Figure \ref{fig_MW_power}(b) shows the [$P_\textrm{a},H$]-contour plot in which the magnetic field is scanned at different power levels at the DUT ($26 - 16$ dBm). A clear decreasing trend of the measured MO signal is seen, until a peak power threshold is reached, at $P_\textrm{a} = 17$ dBm, below which the measured MO signal becomes trivial. Figure \ref{fig_MW_power}(c) plots the integrated amplitude of the MO signal against the square root of the applied power, $\sqrt{P_\textrm{a}}$, and the curve was linearly fitted, confirming the postulated MW power dependence. 

\begin{figure}[htb]
 \centering
 \includegraphics[width=3.3 in]{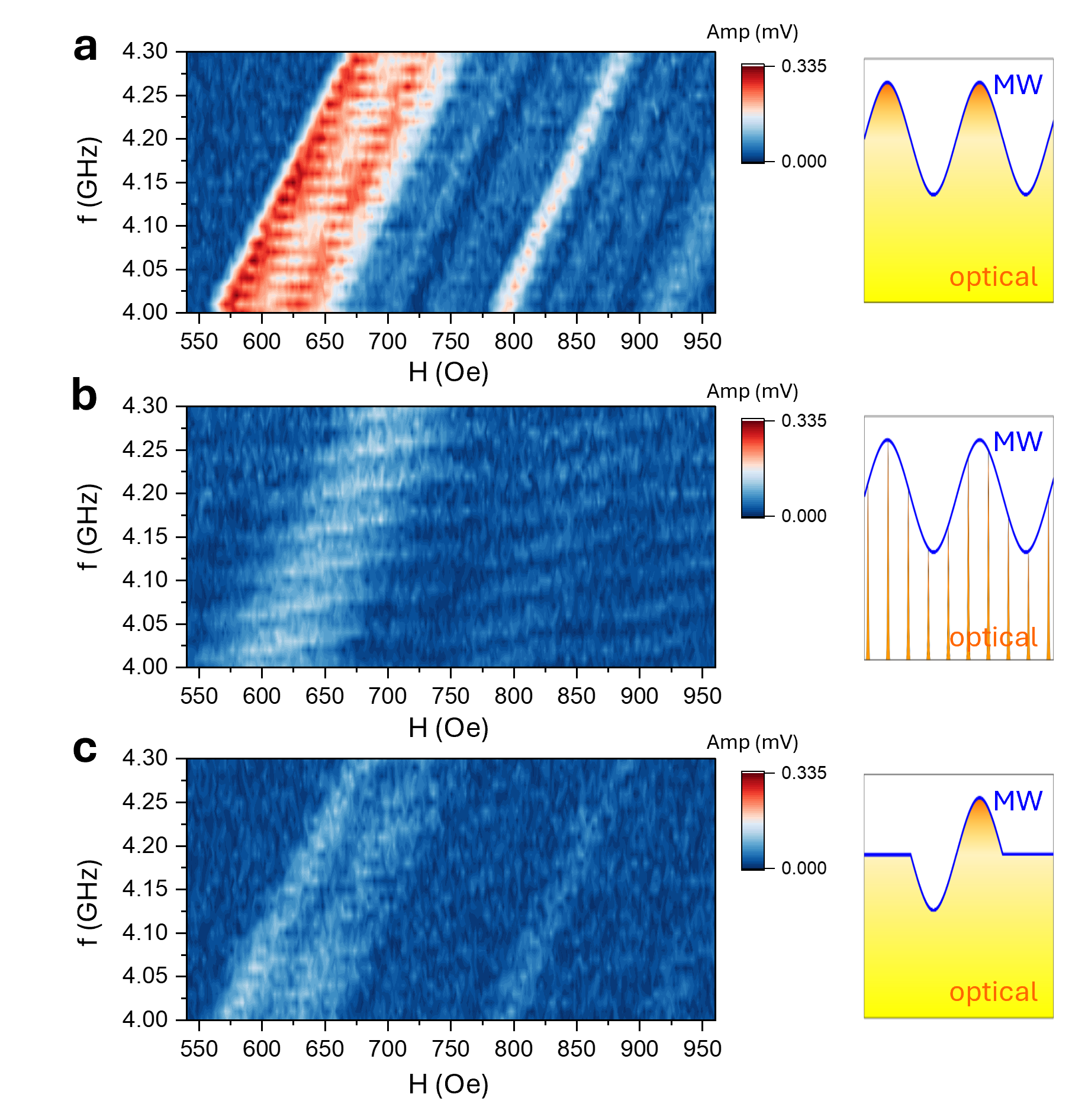}
 \caption{Effect of different combinations of the pump and probe pulse widths. The [$f,H$]-contour plots of magnon spectrum measured at 4 -- 4.3 GHz at different combinations of the MW and optical pulse widths. (a) Optical: 20 ns, MW: 25 ns. (b) Optical: 0.08 ns, MW: 10 ns. (c) Optical: 20 ns, MW: 5 ns.}  
 \label{fig_MW}
\end{figure}

\begin{figure}[htb]
 \centering
 \includegraphics[width=3.6 in]{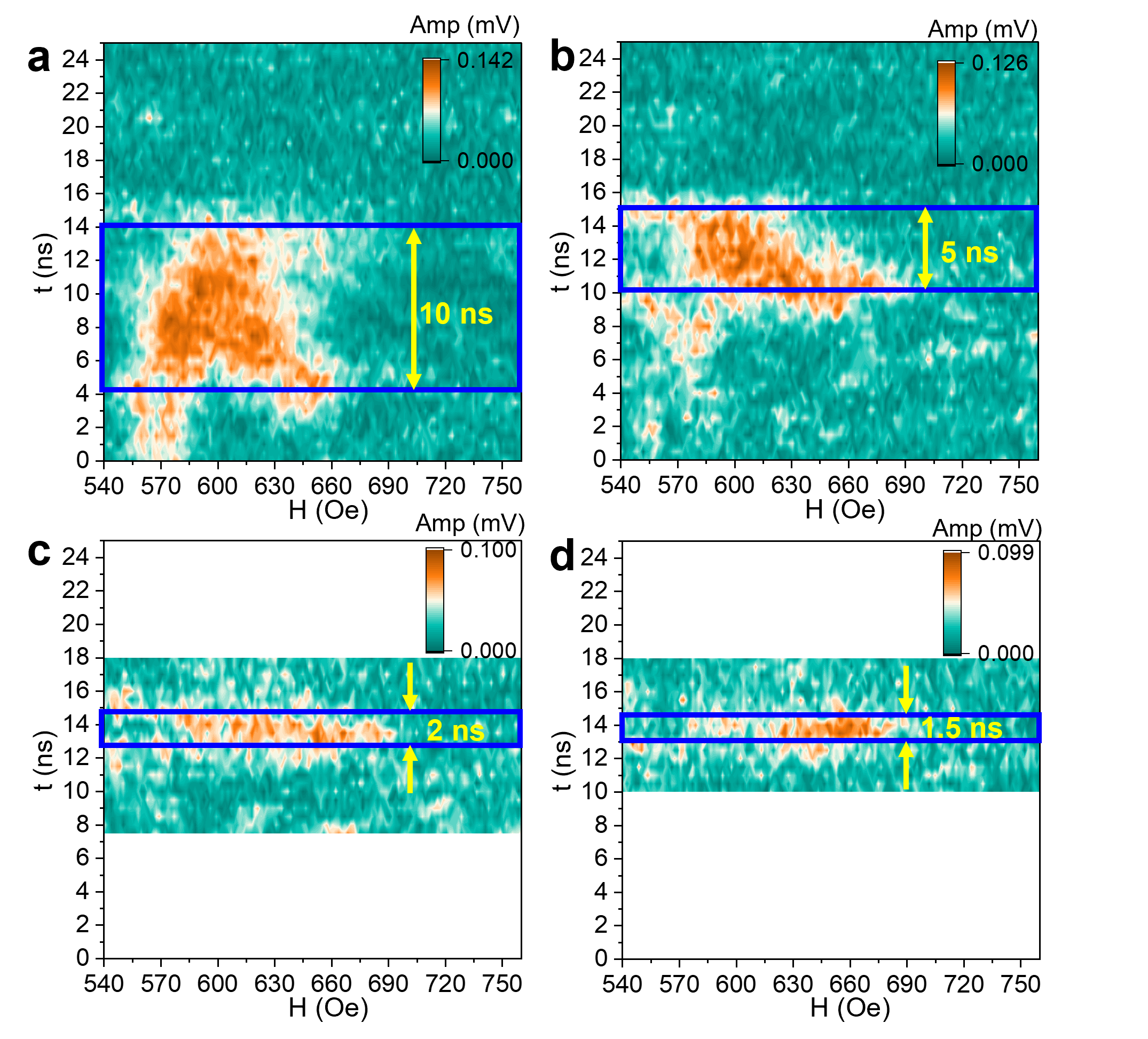}
 \caption{MW pulse mapping. (a-d) Magnon temporal histogram measured by scanning time-delayed optical pulses (80 ps width) across the MW pulse of frequency 4 GHz and various pulse widths: (a) 10, (b) 5, (c) 2, (d) 1.5 (units in ns).  }  
 \label{fig_scan}
\end{figure}

We also investigated the effect of the MW pulse width. Figure \ref{fig_MW} shows the magnon dispersion measured at a frequency range 4 -- 4.3 GHz and at different combinations of MW and optical pulses. Figure \ref{fig_MW}(a) shows the MO signal using both a long MW (25 ns) pulse and a long optical (20 ns) pulse -- both the magnon resonance and the strobe phase alternation (periodic stripe-like fringes) can be clearly observed, whose spectral features resemble very much to the cw scenario \cite{xiong2024phase}. When the optical pulse \textcolor{black}{becomes much shorter} (80 ps) than the MW pulse (10 ns), as in Fig.\ref{fig_MW}(b), the overall MO signal appears weaker due to the low optical probe energy, however, the strobe phase resolution \textcolor{black}{(banding fringes)} is not compromised. This is owing to the generally long optical coherence between light pulses, \textcolor{black}{in spite of their} ultra-short pulse width, and thus the strobe interference can still be well maintained as long as the MW pump pulse is long enough. 

On the other hand, when the MW pulse is significantly reduced (5 ns), in Fig.\ref{fig_MW}(c), the strobe phase contrast substantially diminishes despite using a long enough optical (20 ns) pulse. This is likely due to the decreased sampling rate of the coherent magnon oscillation period (only 5 ns) throughout the 20 ns optical probe pulse. The shorter MW pulses reduced the phase coherence between pulses, and hence the coherence of excited magnon modes, impairing the strobe phase resolution.

\subsection{D. Pump-probe time delays}

The temporal histogram manifesting the time evolution of the non-equilibrium dynamics is the most prominent and unique function of a pump-probe spectroscopy. This is accomplished by incrementally adjusting the probe time positions (time delays) relative to the pump time position. In our hybrid pump-probe measurement, such a time delay can be tuned across an extended time span owing to the combination of the digital(coarse-delay) and optical(fine-delay) components. Besides, in contrast to all-optical pump-probe setups, the MW pump (in our present setup) allows to adjust both the frequency and phase of the excitation signal, so that the temporal evolution of a specific magnon mode (as well as its harmonics) can be traced with amplitude and phase resolutions. 

Figure \ref{fig_scan} shows typical temporal histograms of the strobe light detected PSSW magnon band, obtained via scanning the optical pulse (80 ps) across the MW pump pulse (indicated by the blue enclosures), at selective pulse widths, 10, 5 , 2, and 1.5 ns, see Fig.\ref{fig_scan}(a-d). The excited magnon dynamics faithfully represents the position and width of the MW pump pulse used in each scan. In addition, as the pump pulse width reduces, the excited magnon band broadens, plausibly due to the broadened frequency bandwidth of the pump pulse. The excitation amplitude also scales with the width of the MW pulse, however, it is noted that the strobe light signal is still prominent even at a 1.5-ns MW pulse (in combination with the 80 ps optical probe pulse), in Fig.\ref{fig_scan}(d). With a 7-MHz clock rate used in our experiment, the pump and probe footprints account for only $\sim 1\%$ of each detection cycle.

\subsection{E. Magnon propagation}

\begin{figure*}[htb]
 \centering
 \includegraphics[width=7.0 in]{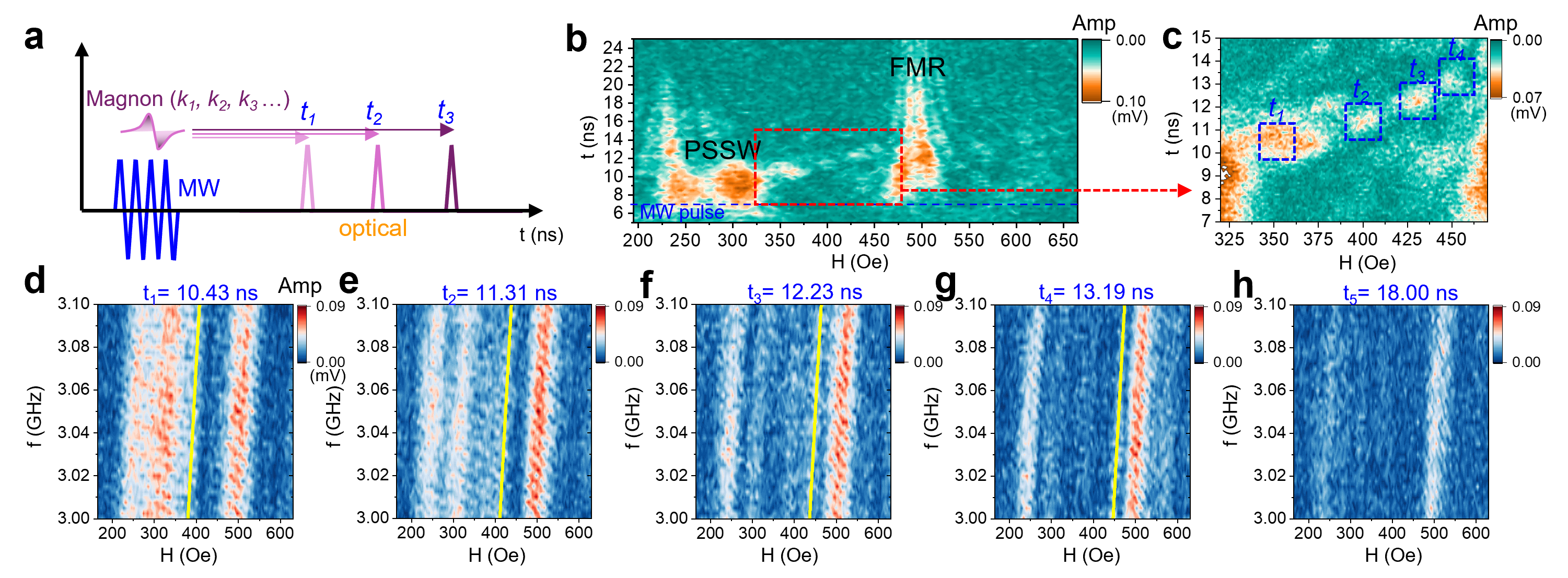}
 \caption{Strobe light detecting magnon propagation. (a) The MW pulse launches a series of PSSW modes ($k_1, k_2, k_3...$) that subsequently travels in-plane with different group velocities and are detected at the optical probe location at different time delays, $t_\textrm{delay}$. (b) Temporal histogram of the magnon excitation, scanned using time-delayed optical probes (80 ps) in the presence of a MW pulse (5 ns, 3 GHz), launched at $t_0$ $\sim$7 ns. The most pronounced excitations are the PSSW magnon band (at $\sim$310 Oe) and the YIG’s FMR mode (at $\sim$500 Oe). (c) Fine-scanned histogram (of the red enclosure in (b)) showing a series of distinct PSSW modes due to propagation. (d-h) The [$f,H$]-contour plots of the magnon dispersion probed at each optical delay corresponding to a distinct PSSW mode, (d) $t_1=10.43$ ns, (e) $t_2=11.31$ ns, (f) $t_3=12.23$ ns, (g) $t_4=13.19$ ns, (h) $t_5=18.00$ ns (away from all PSSW resonances). The yellow curves are theoretical fits to the resonance of each corresponding mode. The mode characteristics are summarized in Table \ref{Tab:vg}.  }  
 \label{fig_travel}
\end{figure*}

Due to the hybrid microwave-optical pump-probe scheme, the pump location \textcolor{black}{(MW signal-line)} and probe location (laser spot) can be spatially offset, allowing to optically probe magnon propagation properties driven by a MW pump pulse. As illustrated in Fig. \ref{fig_travel}(a), a short MW pulse can launch a set of magnon modes in the vicinity of the CPW that are detected by the optical pulses at a distance $D_x$ away from the launching point. Due to their different wavevectors ($k_1, k_2, k_3...$), \textcolor{black}{the magnon modes carry} different group velocities and hence are detected at different time delays ($t_1, t_2, t_3...$). 

Figure \ref{fig_travel}(b) displays the temporal histogram of the magnon excitation, scanned using time-delayed optical probes (80 ps) in the presence of a MW pulse (5 ns, 3 GHz), launched at $t_0 \sim$7 ns. The most pronounced excitations are the PSSW magnon band (at $\sim 310$ Oe) and the YIG's FMR mode (at $\sim 500$ Oe). In addition, a series of distinct PSSW modes in the vicinity can be observed [red enclosure in Fig.\ref{fig_travel}(b)], and are further resolved in a fine-scanned histogram in Fig. \ref{fig_travel}(c). These modes, detected at different time delays ($t_1, t_2, t_3, t_4$), are subject to additional travel due to the finite distance between the CPW signal-line and the probe laser spot. 

To elucidate their wavevectors and group velocities, we fix the optical delay at the specific value corresponding to each PSSW mode ($t_{1\rightarrow4}$), and scan the [$f,H$]-contour plot of the magnon dispersion (with respect to each mode), see Fig.\ref{fig_travel}(d-h). For each time delay, the corresponding PSSW mode is traced by the strobe light with prominent amplitude, apart from the FMR Kittel mode of YIG (that persists for all time delays). Next, we write the explicit expression of the propagating PSSW mode dispersion as a function of the bias magnetic field, $H_y^\textrm{bias}$ \textcolor{black}{(See Appendix for the detailed derivation)}, as:
\begin{equation}
    \begin{aligned} 
    \omega^2 = & \frac{8}{9}\Big(\frac{\gamma K_1}{\mu_0 M_s}\Big)^2+ \\
               & \Big(\gamma H_y^\textrm{bias}+\gamma M_s+\frac{2\gamma A_{ex}}{\mu_0 Ms}(k_x^2+k_z^2) - \frac{\gamma K_1}{\mu_0 M_s}\Big) \cdot \\
               & \Big(\gamma H_y^\textrm{bias}+\frac{2\gamma A_{ex}}{\mu_0 Ms}(k_x^2+k_z^2) \Big). 
    \end{aligned}
\label{Eq:pssw}
\end{equation}
Hence, the group velocity along $x$, $v_{g,x} = \frac{\partial \omega}{\partial k_x}$, is given by: 
\begin{equation}
    \begin{aligned} 
    v_{g,x} = \frac{1}{2\omega} \Bigg( \frac{4\gamma A_{ex}}{\mu_0 Ms} k_x \big(2\gamma H_y^\textrm{bias}+ & \frac{4\gamma A_{ex}}{\mu_0 Ms}(k_x^2+k_z^2)+ \\
              & \gamma M_s - \frac{\gamma K_1}{\mu_0 M_s} \big) \Bigg).   
    \end{aligned}
\label{Eq:vg}
\end{equation}
Given the large YIG thickness, the standing wave component $k_z$ is much smaller than the traveling wave counterpart, $k_x$, at 3 GHz, therefore, $k_x$ and $v_{g,x}$ can be estimated, whose values are summarized in Table \ref{Tab:vg}. \textcolor{black}{The group velocity increases as the wavevector increases, further corroborating the excited PSSW modes in our experiments \cite{gurevich2020magnetization}, whereas for magnetostatic spin-waves (MSSWs), the opposite trend between $k$ and $v_g$ would result. } The extracted values of wavevector and group velocity are in good agreement with previous reports \cite{xiong2020experimental, an2019optimization}. 

Given the estimated group velocities, the distance traveled in a single repetition cycle (142.9 ns) is only tens of $\mu$m, therefore, it is noted that the optically detected magnons could be excited by a MW pulse occurring in prior repetition cycles. As a result, the traveling distance, $D_x$, can be expressed in a general form, as: $D_x \textrm{(nm)} = v_{g,x} \textrm{(m/s)} \cdot (p_i \cdot 142.9 + t_\textrm{delay} + t_\textrm{offset} ) \textrm{(ns)}$, in which the traveling time offset, $t_\textrm{offset}$ is determined by the values of $p_i$, an integer denoting the repetition cycle. Reconciling the four PSSW modes observed in our measurement yields a set of $p_i$ values, $p_{1\rightarrow4}$ = 7, 8, 9, 10, and a traveling time offset $-401.71$ ns. The resultant travel distance $D_x$ is $\sim 0.2$ mm, which is in good agreement with that estimated from the microscope image.

\begin{table}[htb]
\centering
 \begin{tabular}{||c c c c c c||} 
 \hline
 $t_\textrm{delay}$(ns) & $H_y^\textrm{bias}$(Oe) & $p_i$ & $k_x^\textrm{fit}$(m$^{-1}$) & $v_{g,x}$(m/s) & $D_x$($\mu$m) \\ [0.5 ex] 
 \hline\hline
 $t_1$(10.43) & 356.4 & 7 & $1.952 \times 10^7$ & 331.63 & \textbf{202} \\ 
 $t_2$(11.31) & 402.0 & 8 & $1.605 \times 10^7$ & 272.62 & \textbf{205} \\
 $t_3$(12.23) & 432.0 & 9 & $1.328 \times 10^7$ & 225.53 & \textbf{202} \\
 $t_4$(13.19) & 448.5 & 10 & $1.147 \times 10^7$ & 194.84 & \textbf{203} \\ [0.5 ex] 
 \hline
 \end{tabular} 
 \caption{Summary of the delay time($t_\textrm{delay}$), bias field($H_y^\textrm{bias}$), repetition($p_i$), fitted wavevector($k_x^\textrm{fit}$), group velocity($v_{g,x}$), and travel distance($D_x$) associated with the four characteristic propagation PSSW magnon modes. }   
 \label{Tab:vg}
\end{table}

\begin{figure}[htb]
 \centering
 \includegraphics[width=3.0 in]{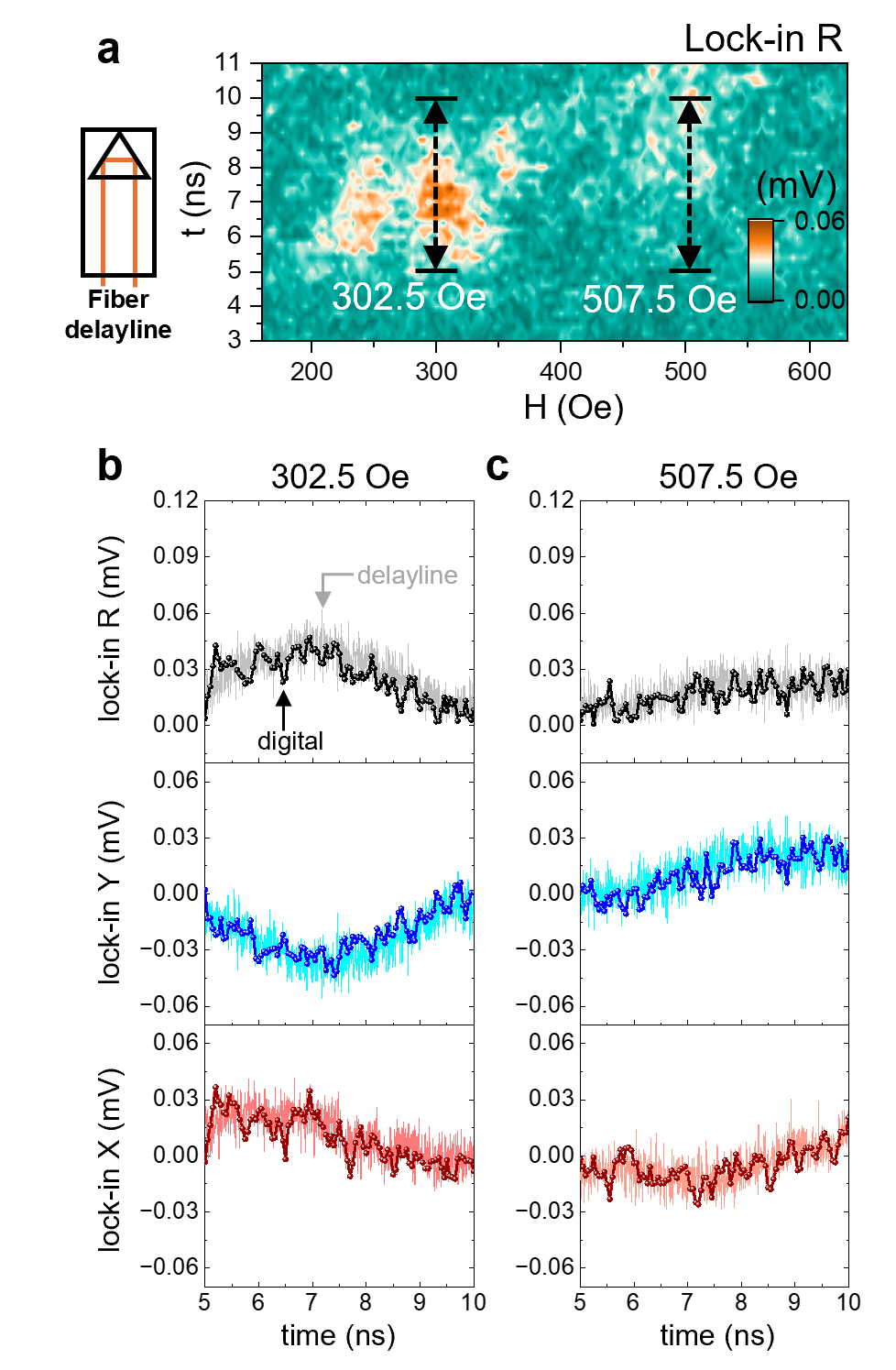}
 \caption{Comparison of digital and optical delay components. (a) Temporal histogram of magnon spectrum, scanned using time-delayed optical probes (80 ps) in the
presence of a MW pulse (2.5 ns, 3 GHz), launched at $t_0 \sim$7 ns. (b,c) lock-in amplifier's \textit{X}, \textit{Y}, and \textit{R} channel signals at selective magnetic fields: (b) $H = 302.5$ Oe, (c) $H = 507.5$ Oe. Symbols: using digital delay (DG645). Curves: using optical fiber delayline. }  
 \label{fig_delayline}
\end{figure}

Last but not least, the co-existing PSSW magnon band and the YIG FMR mode observed in our measurement allows to further elucidate their damping characteristics from their distinct temporal evolutions. As seen in Fig.\ref{fig_travel}(b), while the PSSW excitation sharply ceases once the MW pulse is turned off, the YIG FMR mode, on the other hand, persists for an extended period of time, lasting more than 15 ns beyond the falling edge of the MW pulse.
\textcolor{black}{The observed behavior can be theoretically explained by the mode-dependent damping coefficient, derived from the imaginary part of the complex characteristic frequency (see Appendix for the detailed derivation)}: 
\begin{equation}
    \Gamma^\textrm{PSSW} \approx \frac{1}{2}\alpha \gamma (2H_y^\textrm{bias}+ M_s+\frac{4A_{ex}(k_x^2+k_z^2)}{\mu_0 M_s}),
 \label{Eq:damping}
\end{equation}
\noindent and, the characteristic decay time is its inverse: $\tau_0 = 1/\Gamma^\textrm{PSSW}$, which decreases with increasing wavevector $k$. Thus, the PSSW modes decay faster compared to the FMR mode, due to enhanced, exchange-mediated energy dissipation. The FMR ($k=0$) mode, on the other hand, exhibits the longest lifetime because the third term in the bracket of Eq.\ref{Eq:damping} is zero.

\subsection{F. Reconciling the electrical and optical delays}

Given the hybrid pump-probe setup, the relative delay between the optical pulse (probe) and the MW pulse (pump) can be introduced either electrically (internally from the delay generator), or optically (externally from the fiber delayline). The former operates via electrical signal sampling, and typically functions satisfactorily down to $\sim100$ ps, however, below which the electrical jitter (typically tens of ps) could nontrivially impact the accuracy of the delay time and thus compromises the temporal resolution. On the other hand, the latter functions via manipulating the path of light travel and hence is able to reach a higher resolution, down to $\sim1$ ps, using high-precision translational stages.  

Figure \ref{fig_delayline} compares the results of using electrical and optical delays by scanning a delayed optical pulse (80 ps) across a MW pump pulse (2.5 ns, 3 GHz). First, a rough scan over time, $t$ and magnetic field, $H$ was performed to locate the time position of the MW pulse, i.e. at $t_0 \sim$7 ns, see Fig.\ref{fig_delayline}(a), using the electrical digital delays. Next, we selected two representative field values, one near the PSSW magnon band and the other near the YIG FMR, and scanned the optical pulse from 5 to 10 ns. Figure \ref{fig_delayline}(b) and (c) show the temporal scan traces of the lock-in amplifier's $X,Y,R$ channels, using the electrical delay (symbols) and the optical delay (curves), at $H=302.5$ Oe and 507.5 Oe, respectively. The step size used for the electrical delay is 50 ps, while for the optical delay is 5 ps. The results show good agreement, corroborating the robust strobe light MO detection in our experiment, independent of the delay mechanisms. However, we note that the optical delayline, with its finer step size, can be potentially adopted for resolving ultrafast temporal dynamics, similar to that of a conventional, all-optical pump-probe system.

\section{IV. Summary and Outlook}

In summary, we demonstrate a pump-probe strobe light spectroscopy for sensitive detection of magneto-optical dynamics in a YIG/Py bilayer sample. This technique uses a combinatorial microwave-optical pump-probe setup, leveraging both the high-energy resolution of microwaves and the high-efficiency detection with optical photons. The detected magneto-optical signals strongly depend on the characteristics of both the microwave and the optical pulses, as well as their relative time delays, which can be controlled either digitally(from the common clock) or optically(using delayline). A good magneto-optical sensitivity and coherent stroboscopic character can be maintained even at a microwave pump pulse of 1.5 ns, in combination with an optical probe pulse of 80 ps, under a 7-MHz clock rate, corresponding to a pump-probe footprint of only $\sim 1\%$ in each detection cycle. \textcolor{black}{Further, although a commercial YIG disc with a larger thickness was used in the demonstration, prior reports have indicated that such a technique is in principle applicable to a variety of sample materials and dimensions, such as metallic Py \cite{li2019simultaneous}, CoFe \cite{li2019optical}, and YIG/metal bilayers \cite{xiong2020probing,xiong2020detecting,xiong2022tunable}, and potentially, 2D layered magnets \cite{tang2023spin}. }

The hybrid (microwave-optical) pump-probe scheme allows: \textit{one}, the energy footprints to be significantly reduced in a stroboscopic measurement with high phase-sensitivity; \textit{two}, the free-space portion of the optical layout to be seamlessly integrated with other pump-probe light wavelengths, such as DUV -- VIS systems, to form a multi-color spectroscopy \cite{kirilyuk2010ultrafast,tanksalvala2024element,probst2024unraveling}; \textit{three}, the pumped excitation dynamics to be frequency and phase specific (compared to all-optical schemes); \textit{four}, the time delay to be tuned across an extended time span owing to the combination of digital(coarse-delay) and optical(fine-delay) components; \textit{five}, the pump signal can manifest a more sophisticated format, such as multiplexed frequencies (using rf combiner) and/or timed pulse-train patterns (using pulse combiner), that is often encountered in interferometry experiment involving Rabi \cite{stievater2001rabi,wolz2020introducing,song2025single,fukami2024magnon} and/or Ramsey \cite{clemmen2016ramsey,rani2025high,taylor2008high,makiuchi2024persistent} sequences. The above features make the system promising in the context of hybrid optoelectronic and quantum magnonic systems \cite{yuan2022quantum,xiong2020experimental,xiong2024magnon}, for example, stroboscopic phase tracking under a concurrent pump-probe scheme that is compatible with the superconducting rf platform \cite{li2019strong,hou2019strong,rani2025high,zollitsch2023probing}, and, at the same time, low-power optical probe \cite{karimeddiny2023sagnac,li2019simultaneous,hisatomi2023quantitative,dunagin2025brillouin} for sensing and communicating in cryogenic environments.

\section{Acknowledgements}

The experimental work at UNC-CH was primarily supported by the U.S. Department of Energy (DOE), Office of Science, Basic Energy Sciences under Award Number DE-SC0026305. Part of the work related to the DUT fabrication was supported by the U.S. National Science Foundation (NSF) under award No. OSI-2426642 (W.Z. and B.Y.). Part of the work related to the data curation and modeling was supported by the NSF under award No. DMR-2509513 (W.Z. and J.-M.H). Y.Z. and J.-M.H. also acknowledge the support by the U.S. DOE, Office of Science, Basic Energy Sciences under Award Number DE-SC0020145 as part of the Computational Materials Sciences Program. J.-M.H. acknowledges partial support for manuscript preparation provided by the Wisconsin MRSEC (DMR-2309000). C.T. and W.J. acknowledge support from NSF Grant Nos. DMR-2129879 and DMR-2339615. W.J. acknowledges partial support by U.S. DOE, Office of Science, Grant No. DE-SC0023478. R.S. and D.S. acknowledge financial support from NSF under award No. DMR-2143642.

\section{DATA AVAILABILITY}

The data are available from the authors upon reasonable request.

\section{Appendix}

\renewcommand{\theequation}{A-\arabic{equation}}
\setcounter{equation}{0}  
\renewcommand{\thefigure}{A-\arabic{figure}}
\setcounter{figure}{0}  

\subsection{Derivation of the dispersion relation for the exchange-coupling-dominated magnon mode}

In this Appendix, we analytically derive the  dispersion relation and the mode-dependent damping coefficient for the exchange-coupling-dominated magnon mode based on the linearized Landau–Lifshitz–Gilbert (LLG) equation. The derivations are similar to those reported by Zhuang and Hu \cite{zhuang2022excitation}, except that the Gilbert damping term is kept herein in the linearized LLG equation for studying the $k-$dependent mode damping. For completeness, tensor transformation was performed to account for the (111) orientation of the YIG, which was typically ignored due to the relatively weak magnetocrystalline anisotropy of YIG.

The LLG equation in the YIG layer can be written as, 
\begin{equation}
\frac{\partial{\mathbf{m} }}{\partial t}
=-\gamma (\mathbf{m} \times ( \mathbf{H}^{\mathrm{eff} }+ \mathbf{h})) +
\alpha(\mathbf{m} \times \frac{\partial \mathbf{m} }
{\partial t} )
,
\label{Eq:LLG}
\end{equation}

where $\alpha$ is Gilbert damping coefficient, the effective field $\mathbf{H}^{\mathrm{eff} } = \mathbf{H}^{\mathrm{anis} } + \mathbf{H}^{\mathrm{exch} } + \mathbf{H}^{\mathrm{bias} } + \mathbf{H}^{\mathrm{d} } + \mathbf{h}^{\mathrm{ext} }$, 
$\mathbf{H}^{\mathrm{anis} }$ is the effective magnetocystalline anisotropy field, 
$\mathbf{H}^{\mathrm{exch} }$ is the intralayer exchange coupling field,
$\mathbf{H}^{\mathrm{bias}} = (0, H^{\textrm{bias}}_{y}, 0)$ is the bias magnetic field with $y$ component being nonzero,
$\mathbf{H}^{\mathrm{d}} = (0, 0, -M_{\mathrm{s}}m_{z} )$ is the demagnetization field,
$\mathbf{h}^{\mathrm{ext}} = \mathbf{h}^{0} e^{-\textrm{i} \omega t}$ is the driving MW field in the YIG layer.
As shown in Fig.~\ref{fig_dispersion}(a), one can find that the spin wave is propagating along the $x$ and $z$ directions.
The magnetization $\mathbf{m}$ can be written as, 
$\mathbf{m} = \mathbf{m}^{0} + \Delta \mathbf{m}^{0} e^{\mathrm{i} (k_zz+k_xx-\omega t)}$,
where $\mathbf{m}^{0} = (m_x^0, m_y^0, m_z^0) = (0,1,0)$ is the static magnetization under the bias magnetic field $\mathbf{H}^{\mathrm{bias}}$, 
and $\Delta \mathbf{m}^{0} = (\Delta m_x^0, \Delta m_y^0, \Delta m_z^0)$ is the amplitude of the magnetization oscillation.

Neglecting the higher-order terms, the magnetocrystalline energy density of the YIG layer can be written as,
\begin{equation}
f^{\textrm{anis}} = K_1 (m_1^2 m_2^2 + m_2^2 m_3^2 + m_1^2 m_3^2)
,
\label{Eq:f_anis}
\end{equation}
where the subscripts $1,2,3$ represent the axes in the crystal physics coordinate system of the cubic phase, where $x_1 // [100]_\mathrm{c}$, $x_2 // [010]_\mathrm{c}$, $x_3 // [001]_\mathrm{c}$. 
In a (111)-oriented YIG disc, one has $z // [111]_\mathrm{c}$, $x // [\bar{1}10]_\mathrm{c}$, $y // [\bar{1} \bar{1}2]_\mathrm{c}$, and the relationship between the magnetization in the crystal physics coordinate ($m_1,m_2,m_3$) and that in the lab coordinate ($m_x,m_y,m_z$) can be written as,  

\begin{equation}
\begin{pmatrix}
m_1 \\
m_2 \\
m_3
\end{pmatrix}
=
\begin{pmatrix}
\frac{1}{\sqrt{2}} & \frac{1}{\sqrt{6}} & \frac{1}{\sqrt{3}} \\
-\frac{1}{\sqrt{2}} & \frac{1}{\sqrt{6}} & \frac{1}{\sqrt{3}} \\
0 & -\sqrt{\frac{2}{3}} & \frac{1}{\sqrt{3}}
\end{pmatrix}
\begin{pmatrix}
m_x \\
m_y \\
m_z
\end{pmatrix}
\label{Eq:cry2lab}
\end{equation}

Substituting the Eq.~\ref{Eq:cry2lab} into Eq.~\ref{Eq:f_anis}, one can get,

\begin{equation}
    \begin{aligned} 
f^{\textrm{anis}} = &K_1 (\frac{1}{4}m_x^4 + \frac{1}{4} m_y^4 + \frac{1}{3} m_z^4 +\frac{1}{2} m_x^2m_y^2 \\
&- \sqrt{2}m_x^2m_ym_z + \frac{\sqrt{2}}{3} m_y^3m_z),
\end{aligned}
\label{Eq:f_anis_lab}
\end{equation}

The magnetocystalline anisotropy effective field, $\textbf{H}^\textrm{anis} = -\frac{1}{\mu_0 M_{\textrm{s}}} \frac{\partial f^{\textrm{anis}}}{\partial \textbf{m}}$, can be written as,
\begin{equation}
\begin{aligned} 
H^{\textrm{anis}}_x  &= -\frac{K_1m_x}{\mu_0 M_{\textrm{s}}} ( m_x^2 + m_y^2 - 2 \sqrt{2} m_y m_z), \\
H^{\textrm{anis}}_y &= -\frac{K_1}{\mu_0 M_{\textrm{s}}} ( m_x^2m_y + m_y^3 - \sqrt{2} m_x^2 m_z + \sqrt{2}m_y^2 m_z), \\
H^{\textrm{anis}}_z &= -\frac{K_1}{3\mu_0 M_{\textrm{s}}} ( 3 \sqrt{2}m_x^2m_y -\sqrt{2} m_y^3 - 4 m_z^3),
\end{aligned} 
\label{Eq:H_anis}
\end{equation}

The intralayer exchange coupling field $\textbf{H}^\textrm{exch}$ can be written as,
\begin{equation}
\begin{aligned} 
\textbf{H}^{\textrm{exch}}  = \frac{2A_{\textrm{ex}}}{\mu_0 M_{\textrm{s}}} \nabla^2 \textbf{m} =
\frac{2A_{\textrm{ex}}}{\mu_0 M_{\textrm{s}}}
\begin{pmatrix}
\frac{\partial^2 m_x}{\partial x^2} +  \frac{\partial^2 m_x}{\partial z^2}\\
\frac{\partial^2 m_y}{\partial x^2} +  \frac{\partial^2 m_y}{\partial z^2}\\
\frac{\partial^2 m_z}{\partial x^2} +  \frac{\partial^2 m_z}{\partial z^2}
\end{pmatrix}
,
\end{aligned}
\label{Eq:H_exch}
\end{equation}

Substituting $\mathbf{m} = \mathbf{m}^{0} + \Delta \mathbf{m}^{0} e^{\mathrm{i} (k_zz+k_xx-\omega t)}$ into Eq.~\ref{Eq:H_anis} and Eq.~\ref{Eq:H_exch}, Eq.~\ref{Eq:LLG} can be rewritten into the following form with the $k-$dependent intrinsic susceptibility tensor $\boldsymbol{\chi}$, 

\begin{equation}
\begin{aligned} 
\begin{pmatrix}
\Delta m_x^0 e^{-\textrm{i}\omega t}\\
\Delta m_y^0 e^{-\textrm{i}\omega t}\\
\Delta m_z^0 e^{-\textrm{i}\omega t}
\end{pmatrix}
=
\boldsymbol{\chi}
\begin{pmatrix}
h_x^0 e^{-\textrm{i}\omega t}\\
h_y^0 e^{-\textrm{i}\omega t}\\
h_z^0 e^{-\textrm{i}\omega t}
\end{pmatrix}&
,\\
\boldsymbol{\chi} = \gamma \boldsymbol{\Omega}^{-1}
\left(
\begin{array}{ccc}
0 & -1 & 1 \\
1 & 0 & -1 \\
-1 & 1 & 0 \\
\end{array}
\right)&\\
= \gamma 
\left(
\begin{array}{ccc}
i\omega & A_{xy} & A_{xz} \\
A_{yx} & i\omega & 0 \\
A_{zx} & 0 & i\omega \\
\end{array}
\right)^{-1}&
\left(
\begin{array}{ccc}
0 & -1 & 1 \\
1 & 0 & -1 \\
-1 & 1 & 0 \\
\end{array}
\right),\\
A_{xy} = \frac{4 \sqrt{2} \gamma K_1}{3 \mu_0 M_{\textrm{s}}}&, \\
A_{yx} = -\frac{\sqrt{2} \gamma K_1}{3 \mu_0 M_{\textrm{s}}}&, \\
A_{xz} = \gamma (H_y^{\textrm{bias}}+M_{\textrm{s}})+\frac{2 \gamma A_{\textrm{ex}}}{\mu_0 M_{\textrm{s}}}(k_x^2+k_z^2)& -\frac{\gamma K_1}{\mu_0 M_{\textrm{s}}} +\textrm{i} \alpha \omega, \\
A_{zx} = -\gamma H_y^{\textrm{bias}} - \frac{2 \gamma A_{\textrm{ex}}}{\mu_0 M_{\textrm{s}}}(k_x^2&+k_z^2) -\textrm{i} \alpha \omega, 
\end{aligned}
\label{Eq:susceptibility}
\end{equation}

At the resonant frequency of an exchange-coupling-dominated magnon mode (including the PSSW mode) or a Kittel mode under a specific bias magnetic field, the modulus of the susceptibility tensor $\boldsymbol{\chi}$ shown above maximizes. Such a resonance condition corresponds to the pole of the susceptibility tensor $\boldsymbol{\chi}$, at which the determinant of the matrix $\boldsymbol{\Omega}$ is zero. In general, the eigenfrequency is complex-valued, and the resonance is satisfied when the real part of the determinant is zero. Thus, by solving $\mathrm{Re}(\det \boldsymbol{\Omega}) = 0$, one can derive the expression of the complex-valued resonance frequency of the exchange-coupling-dominated magnon mode, given by,

\begin{equation}
\omega = \omega^{0} + \textrm{i} \Gamma^{0} = \sqrt{-A_{xy}A_{yx}-A_{xz}A_{zx}},
\label{Eq:omega_rensonance}
\end{equation}
where the real part $\omega^{0}$ represents the resonant frequency of the exchange-coupling-dominated magnon mode, while the imaginary part $\Gamma^{0}$ describes its damping coefficient. Specifically, one has,

\begin{equation}
\omega^{0} = \sqrt{
\begin{aligned} 
    & \frac{8}{9}\Big(\frac{\gamma K_1}{\mu_0 M_s}\Big)^2+ \Big(\gamma H_y^\textrm{bias}+\frac{2\gamma A_{ex}}{\mu_0 Ms}(k_x^2+k_z^2) \Big) \cdot\\
               & \Big(\gamma H_y^\textrm{bias}+\gamma M_s+\frac{2\gamma A_{ex}}{\mu_0 Ms}(k_x^2+k_z^2) - \frac{\gamma K_1}{\mu_0 M_s}\Big) 
\end{aligned}
}
\label{Eq:omega_Re}
\end{equation}

If letting $k_x = k_y=0$, Eq. (A-9) reduces to the resonance frequency of the Kittel mode magnon in (111) YIG. When the anisotropy effective field is neglected, i.e. $K_1 = 0$, the damping coefficient $\Gamma^{0}$ can be approximately expressed as,
\begin{equation}
\Gamma^{0} =\frac{1}{2} \alpha \gamma (2 H_y^{\textrm{bias}}+M_{\textrm{s}} + 
\frac{4 A_{\textrm{ex}}(k_x^2 + k_z^2)}{\mu_0 M_{\textrm{s}}}). 
\label{Eq:omega_Im}
\end{equation}

\bibliography{sample_ppstrobe}

\end{document}